\documentclass[10pt]{article}
\usepackage[utf8]{inputenc}
\usepackage[T1]{fontenc}
\usepackage{microtype}
\usepackage[autostyle=true, style=english]{csquotes}
\usepackage{geometry}
\usepackage{graphicx}
\usepackage{caption}
\usepackage{subcaption}
\usepackage{float}
\usepackage{tikz}
\usepackage{chemformula}
\usepackage{enumitem}
\usepackage{authblk}
\usepackage[colorlinks=true, urlcolor=blue, linkcolor=black, citecolor=black]{hyperref}
\usepackage{xcolor}

\begin{document}

\sloppy

\title{MaterialsGalaxy: A Platform Fusing Experimental and Theoretical Data in Condensed Matter Physics}

\date{}

\author[1,2]{Tiannian Zhu}
\author[1,2]{Zhong Fang}
\author[1,2]{Quansheng Wu\thanks{quansheng.wu@iphy.ac.cn}}
\author[1,2]{Hongming Weng\thanks{hmweng@iphy.ac.cn}}
\affil[1]{%
Beijing National Laboratory for Condensed Matter Physics and Institute of Physics,
Chinese Academy of Sciences, Beijing 100190, China
}
\affil[2]{%
University of Chinese Academy of Sciences, Beijing 100049, China
}

\maketitle

\section*{Abstract}
\noindent
Modern materials science generates vast and diverse datasets from both experiments and computations, yet these multi--source, heterogeneous data often remain disconnected in isolated \enquote{silos}. Here, we introduce MaterialsGalaxy, a comprehensive platform that deeply fuses experimental and theoretical data in condensed matter physics. Its core innovation is a structure similarity-driven data fusion mechanism that quantitatively links cross-modal records—spanning diffraction, crystal growth, computations, and literature—based on their underlying atomic structures. The platform integrates artificial intelligence (AI) tools, including large language models (LLMs) for knowledge extraction, generative models for crystal structure prediction, and machine learning property predictors, to enhance data interpretation and accelerate materials discovery. We demonstrate that MaterialsGalaxy effectively integrates these disparate data sources, uncovering hidden correlations and guiding the design of novel materials. By bridging the long-standing gap between experiment and theory, MaterialsGalaxy provides a new paradigm for data-driven materials research and accelerates the discovery of advanced materials.

\noindent
\textbf{Keywords:} MaterialsGalaxy, Data fusion, Materials gene, Materials database

\noindent
\textbf{PACS:} 07.05.Mh, 61.50.Ah, 71.15.-m, 61.05.cc

\section{Introduction}
\noindent
The fields of condensed matter physics and materials science are undergoing a profound, data-intensive transformation. Decades of experimental exploration and theoretical computation have amassed invaluable data, from experimentally determined crystal structures (e.g., ICSD\cite{zagorac_recent_2019}, COD\cite{grazulis_crystallography_2009,grazulis_crystallography_2012}, CSD\cite{groom_cambridge_2016}, and the Pauling File\cite{villars_pauling_2004}) to properties derived from first-principles calculations (e.g., in databases like Materials Project\cite{jain_commentary_2013}, AFLOW\cite{curtarolo_aflow_2012}, OQMD\cite{kirklin_open_2015}, MatCloud\cite{yang_matcloud_2018}, NOMAD\cite{draxl_nomad_2019}, Materials Cloud\cite{talirz_materials_2020}, Atomly\cite{miao_atomly}). This data wealth, coupled with the rise of the data-driven \enquote{fourth paradigm}\cite{hey_fourth_2009}, offers unprecedented opportunities for materials discovery\cite{agrawal_perspective_2016,ramprasad_machine_2017,lookman_active_2019,schleder_dft_2019,merchant_scaling_2023,butler_machine_2018,choudhary_recent_2022}. It enables the systematic analysis of massive datasets to uncover and interpret hidden structure-property relationships\cite{zhong_explainable_2022,vu_understanding_2023}, accelerate the rational design of novel materials\cite{sanchez-lengeling_inverse_2018,gubernatis_machine_2018,ma_mlmd_2024}, and predict their performance with increasing accuracy\cite{oganov_structure_2019,chan_application_2022,griesemer_accelerating_2023}. Indeed, the synergistic integration of artificial intelligence, high-performance computing, and automated experimentation is emerging as a powerful strategy to enrich and accelerate every stage of the discovery cycle\cite{stein_progress_2019,pyzer-knapp_accelerating_2022,szymanski_autonomous_2023}.

Despite this abundance of data from structured databases, the full potential of these data resources remains largely untapped due to the pervasive \enquote{data silo} phenomenon. An even larger reservoir of knowledge, including crucial synthesis details, resides within the unstructured text of scientific literature. Experimental data, theoretical calculations, and literature-extracted knowledge are fundamentally disparate, differing in formats, naming conventions, precision standards, and acquisition methods. This inherent heterogeneity, compounded by a lack of standardized interoperability protocols, severely hinders cross-source integration and analysis\cite{kalidindi_materials_2015,himanen_datadriven_2019}. Consequently, researchers face significant challenges in comparing and integrating data from these distinct origins, a bottleneck that diminishes research efficiency.

To address this, the materials science community has initiated crucial data integration efforts. The OPTIMADE consortium\cite{andersen_optimade_2021}, for example, has made significant strides in providing a unified Application Programming Interface (API) for computational materials databases, enhancing interoperability among them. Concurrently, specialized NLP models have been developed to extract information from scientific literature\cite{tshitoyan_unsupervised_2019,gupta_matscibert_2022,pyzer-knapp_foundation_2025,jiang_applications_2025}. However, such efforts predominantly focus on homogeneous data sources (e.g., linking computational databases). The deep, cross-modal fusion of experimental data with theoretical computations—a far more complex challenge due to fundamental differences in data generation, semantics, and precision—remains a critical and largely unsolved frontier.

Bridging the divide between experimental and theoretical data holds immense scientific merit. High-quality experimental data provide the ground truth for validating and refining computational models\cite{green_fulfilling_2017,wu_universal_2024}. Crucially, even experimental failures or \enquote{negative results} are invaluable, as they provide critical constraints that help define the boundaries of successful synthesis or desired properties, thereby further refining predictive models\cite{raccuglia_machinelearningassisted_2016}. Conversely, theoretical calculations offer predictive guidance for exploring materials yet to be synthesized or characterized\cite{zhang_topological_2009,weng_weyl_2015}. An effective fusion of these two data modalities would establish a powerful closed loop, where experiment and theory mutually validate and accelerate one another, fostering more accurate models and hastening the discovery of new materials.

To address these challenges, we developed the MaterialsGalaxy platform, designed for the deep fusion of heterogeneous experimental and theoretical databases in condensed matter physics. Our core innovation is a data-linking methodology centered on crystal structure similarity. We transform crystal structures into fixed-length numerical vectors, or \enquote{fingerprints}, that encode key chemical and structural features. While advanced, end-to-end representation learning methods like graph neural networks (GNNs) and continuous-filter convolutional networks offer state-of-the-art performance\cite{xie_crystal_2018,chen_graph_2019,schutt_schnet_2018}, for this foundational work, we opted for a robust and interpretable feature engineering approach from the matminer library\cite{ward_matminer_2018}. These descriptor-based fingerprints can be generated orders of magnitude faster than deep learning embeddings, are deterministic, and their components (e.g., mean atomic radius, packing efficiency) have clear physical and chemical meaning. These vectors are then efficiently indexed in a vector database. Leveraging this index, we perform high-speed similarity searches to dynamically link disparate data records—from experimental synthesis to theoretical properties—that correspond to the same or similar materials, effectively dismantling data silos. Beyond this central fusion engine, MaterialsGalaxy integrates a synergistic suite of AI tools, including a domain-specific large language model TopoChat\cite{xu_enhancing_2024}, a generative model for crystal structure prediction (Con-CDVAE)\cite{ye_concdvae_2024}, and machine learning models for property prediction. These tools, coupled with the fused data, create a powerful ecosystem for intelligent data analysis and accelerated materials discovery.

This paper elucidates the architecture and data fusion methodology of the MaterialsGalaxy platform. We demonstrate its capabilities through application examples that connect experimental and theoretical data to facilitate materials discovery. Finally, we discuss the broader implications of this approach for the research paradigm in condensed matter physics and materials science, aiming to provide a unified data infrastructure that fosters deeper synergy between experiment and theory.

\section{Results}

\subsection{Platform architecture}
\begin{figure*}[t!]
  \centering
  \includegraphics[width=1.0\textwidth]{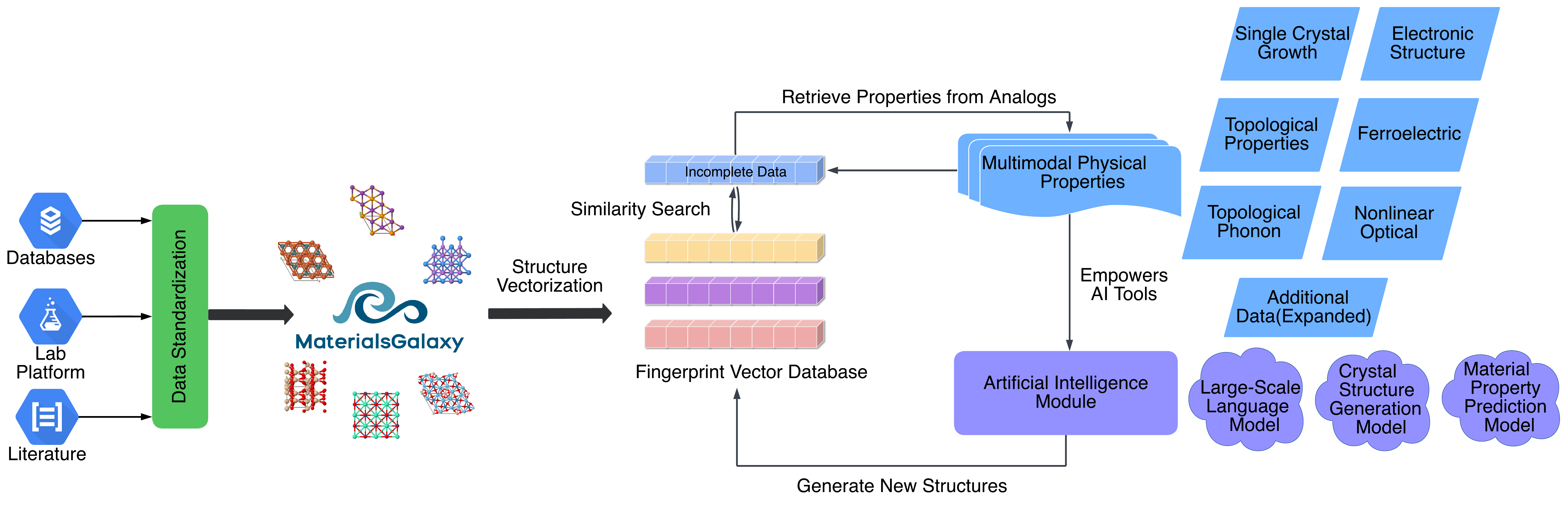}
  \caption{
    \textbf{Architecture of the MaterialsGalaxy platform.} The platform employs a systematic workflow to fuse heterogeneous data from three primary channels: (1) existing public databases, (2) electronic laboratory notebooks, and (3) automated literature extraction. Raw data first undergo a rigorous standardization process. The core innovation is the structure vectorization module, which uses representation learning to generate a unique fingerprint for each crystal structure. These fingerprints are indexed in a high-performance vector database, enabling a similarity matching engine to dynamically link disparate records. The resulting fused data backbone supports a rich application layer featuring interactive querying, visualization tools, a RESTful API, and a suite of integrated AI tools (e.g., LLM-based assistants, generative models, and property predictors). Crucially, this architecture not only connects siloed experimental and theoretical data but also enriches them, creating a comprehensive, multi-modal profile for each material based on shared structural features.
  }
  \label{platform_architecture}
\end{figure*}

\noindent
The overall architecture of the MaterialsGalaxy platform, illustrated in Fig. \ref{platform_architecture}, is engineered to systematically address the challenge of heterogeneous data integration in materials science. At its core, the architecture follows a multi-stage workflow designed for robust data processing and intelligent analysis. It begins with a Data Acquisition and Standardization Layer that ingests and harmonizes multimodal data from disparate sources, including public databases, electronic lab notebooks, and the scientific literature.

The central component is a Structure-driven Fusion Engine. This engine leverages representation learning to vectorize crystal structures and employs a vector database for high-speed similarity matching. This mechanism is the key to linking otherwise disconnected records. Critically, this fusion process creates a holistic, multi-modal profile for each material. For instance, an experimental record, which might originally contain only synthesis conditions and a diffraction pattern, can be dynamically linked to its theoretical counterparts in the fused database, instantly augmenting it with computed properties like band structure, formation energy, and topological invariants. This cross-modal enrichment allows researchers to rapidly gain a comprehensive understanding of a material's properties, seamlessly bridging the gap between its experimental realization and theoretical characteristics.

This fused data backbone supports a versatile Application and Analysis Layer, which provides user-facing functionalities such as interactive querying, data visualization, API access, and a suite of integrated AI tools for property prediction and materials discovery. This cohesive design establishes a systematic solution to not only bridge data silos but also to create a synergistically enriched data ecosystem that empowers advanced, data-driven research.

\subsection{Data Sources and Integration Strategy}
\noindent
The MaterialsGalaxy platform is built upon a diverse and growing collection of multi-source, heterogeneous data from the field of condensed matter physics, hosted at Condensed Matter Physics Data Center, Institute of Physics, Chinese Academy of Sciences. Our data acquisition strategy follows a three-pronged approach: (1) aggregation of established public databases, (2) automated ingestion from our in-house electronic laboratory platform (MatElab)\cite{iphy_matelab}, which captures curated experimental records with full provenance, and (3) automated extraction of structural and property data from the scientific literature. This multi-channel approach ensures the construction of a comprehensive data ecosystem spanning material synthesis, characterization, and theoretical properties.

\begin{figure*}[t!]
  \centering
  \begin{subfigure}[t]{0.56\textwidth}
    \centering
    \begin{tikzpicture}[baseline=(image.north)]
      \node[anchor=south west] (image) at (0,0) {\includegraphics[width=0.9\textwidth]{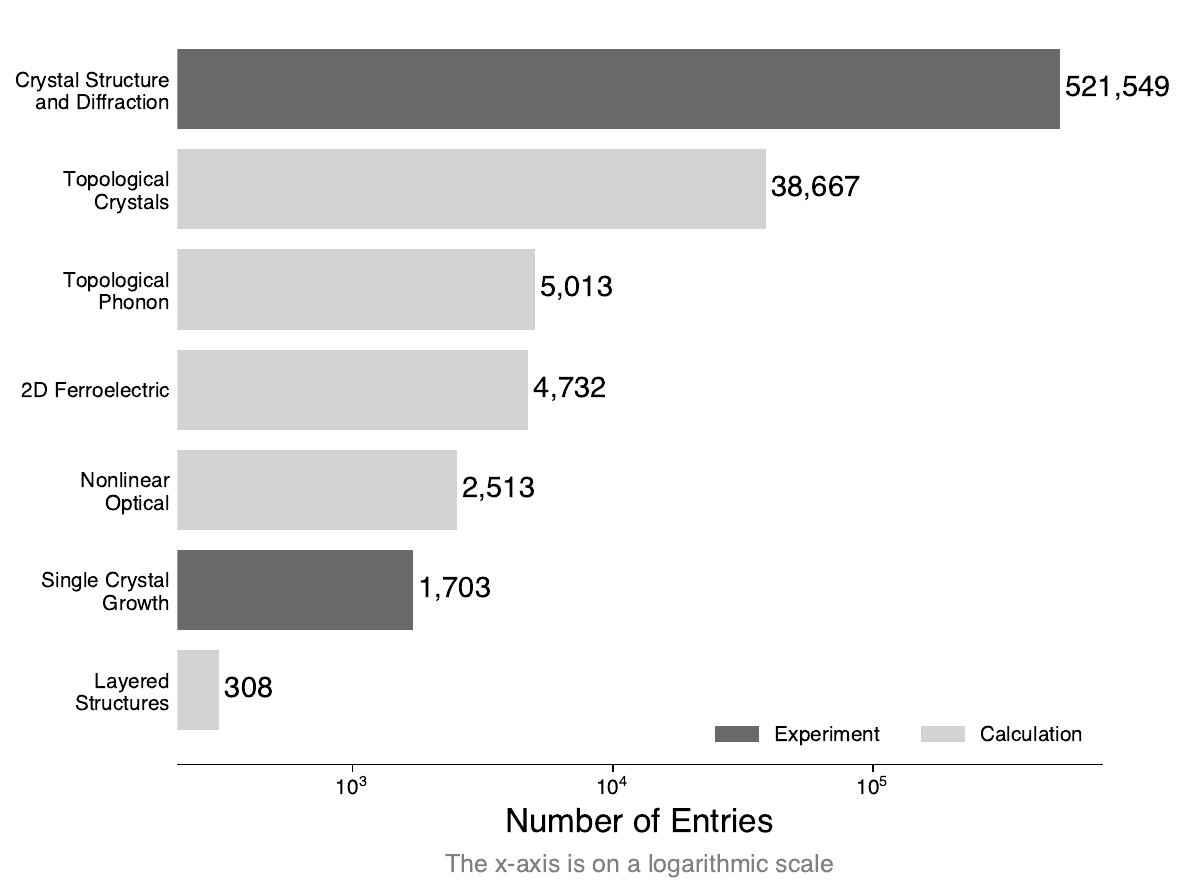}};
      \node[anchor=north east] at (image.north west) {\textbf{a}};
    \end{tikzpicture}
  \end{subfigure}
  \hfill
  \begin{subfigure}[t]{0.43\textwidth}
    \centering
    \begin{tikzpicture}[baseline=(image.north)]
      \node[anchor=south west] (image) at (0,0) {\includegraphics[width=0.89\textwidth]{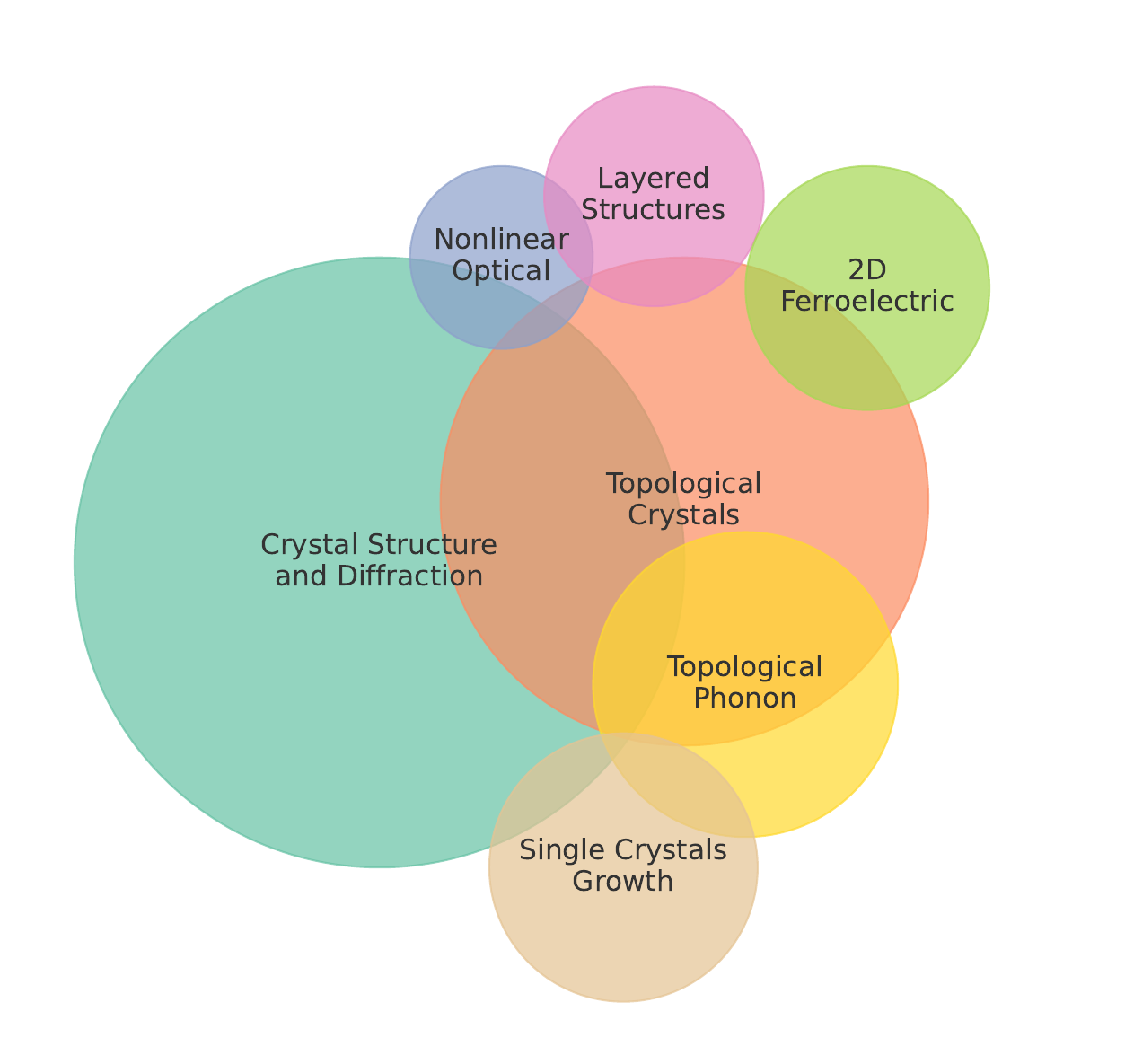}};
      \node[anchor=north east] at (image.north west) {\textbf{b}};
    \end{tikzpicture}
  \end{subfigure}
  \caption{\textbf{Overview of integrated data sources and their heterogeneity.} \textbf{a} Distribution of entries across the primary integrated databases, with experimental sources shown in blue and theoretical/computational sources in orange. The y-axis is on a logarithmic scale to accommodate the wide range of data volumes. The collection includes a large experimental crystal structure database (COD-derived), various computational property databases (e.g., topological materials, topological phonons), and a unique database of single-crystal growth experiments. (b) A conceptual Venn diagram illustrating the complex relationships of overlap and uniqueness among different data modalities. This highlights the core challenge of data heterogeneity, where, for instance, the materials space of experimental synthesis records, theoretically predicted topological materials, and the general crystal structure database are partially intersecting yet distinct, necessitating a robust data fusion strategy.}
  \label{fig:source_collection}
\end{figure*}

The platform currently integrates a wide spectrum of data modalities, including crystal structures, electronic band structures, topological classifications, phonon dispersions, ferroelectric properties, non-linear optical responses, and single-crystal growth recipes. The scale and diversity of these data sources are quantitatively summarized in Fig. \ref{fig:source_collection}a. The experimental cornerstone of our collection is a large-scale crystal structure database\cite{cmpdc_diffraction} comprising nearly 500,000 entries derived from the Crystallography Open Database (COD)\cite{grazulis_crystallography_2009,grazulis_crystallography_2012} and supplemented by literature mining. This is complemented by a unique single-crystal growth database sourced from our MatElab platform\cite{iphy_matelab}, which contains ~2,000 detailed experimental records documenting synthesis parameters and characterization results.

On the theoretical front, MaterialsGalaxy incorporates several high-throughput computation databases. Key among these is a comprehensive topological materials database\cite{cmpdc_materiae}, containing over 8,000 unique materials identified as topologically non-trivial (e.g., topological insulators, semimetals) from a screening of more than 28,000 candidates\cite{zhang_catalogue_2019}. Similarly, a topological phonon database provides phononic band structures and classifications for over 5,000 materials\cite{li_computation_2021,synl_phonon}. Additional computational datasets cover properties such as 2D ferroelectricity\cite{cmpdc_ferroelectric} and non-linear optical coefficients\cite{cmpdc_nlo}.

The primary challenge addressed by our platform stems from the inherent heterogeneity and fragmentation of these data sources. As conceptually illustrated by the Venn diagram in Fig. \ref{fig:source_collection}b, these datasets exhibit complex overlaps and complementarities. For example, the broad chemical space of experimental crystal structures (including organics) contrasts with the inorganic focus of most computational databases. Different property databases may share materials but describe orthogonal physical phenomena. The experimental growth database provides unique synthesis context that is often decoupled from theoretical entries. This intricate landscape of data types, formats, precision levels, and semantic contexts necessitates the systematic standardization and fusion methodologies detailed in the following sections.

\subsection{Data Standardization}
\noindent
A rigorous and automated data standardization pipeline is the bedrock of the MaterialsGalaxy platform, transforming raw, heterogeneous inputs into a consistent, analysis-ready format. This initial step is critical for constructing a truly \enquote{AI-ready} dataset. By this, we mean that the data is not only clean and structured but is also semantically rich and primed for effective use by machine learning algorithms, thus preventing issues like data leakage or biased model training that arise from inconsistent inputs. This disciplined approach is essential, as disparities in crystallographic conventions, data formats, and physical units across sources would otherwise prevent reliable data fusion and undermine model performance.

The standardization process is centered on establishing a canonical representation for crystal structures, the universal anchors for data linking. All incoming structural data are processed using the pymatgen\cite{ong_python_2013} and spglib\cite{togo_spglib_2024} libraries to parse, validate, and resolve symmetries according to IUCr conventions, ensuring identical materials map to a single, unique representation. For associated property data, we developed a formal data schema that enforces standardized nomenclature and units. Crucially, this schema was not developed in isolation; it is the product of close collaboration with domain experts—both the original data producers and active researchers—ensuring that our standards reflect the nuanced requirements and best practices of the condensed matter physics community. This schema is programmatically enforced using data validation libraries like Pydantic, guaranteeing that every data point adheres to a predefined type and structure before ingestion. This is particularly crucial for parsing the output of diverse first-principles calculation packages, such as the widely used Vienna Ab initio Simulation Package (VASP)\cite{kresse_efficient_1996}, ensuring that computational parameters are captured consistently. This systematic process captures not only the data values but also essential metadata (e.g., computational parameters or experimental conditions), providing the robust and reliable foundation required for the structure-based data fusion mechanism described next.

\subsection{Core data fusion: Structure-based similarity linking}
\noindent

Having established a foundation of standardized, canonical data, we implement our core innovation: a dynamic data fusion mechanism driven by crystal structure similarity. This approach directly tackles the long-standing challenge of linking records across heterogeneous databases. Traditional methods for identifying similar crystal structures, such as the structure matching algorithms found in libraries like Pymatgen\cite{ong_python_2013}, rely on direct, pairwise comparisons of atomic coordinates and cell parameters. While precise for near-identical structures, these methods suffer from two critical drawbacks for large-scale data fusion: they are computationally expensive, often scaling poorly to millions of comparisons, and they are brittle, struggling to identify structurally related but not identical phases (e.g., those with minor distortions or different elemental decorations). Our approach overcomes these limitations by moving from direct comparison to a highly efficient, vector-space similarity search.

Our fusion workflow comprises two key stages. First, in an offline pre-processing step, every standardized crystal structure is transformed into a fixed-length numerical vector, or structural fingerprint, using a representation learning algorithm. For this, we employ the SiteStatsFingerprint featurizer from the matminer library\cite{ward_matminer_2018}, which encodes rich information about local atomic environments into a high-dimensional vector. This process effectively maps the complex, variable-sized crystal graph into a unified, machine-readable vector space where geometric and chemical similarity are represented by proximity.

Second, and most critically, data fusion occurs dynamically at query time through a high-performance vector search index. All structural fingerprints are indexed using Approximate Nearest Neighbor (ANN) algorithms (e.g., HNSW-based graphs\cite{malkov_efficient_2020}), enabling sub-second similarity searches. When a user views a specific material entry, the platform performs multiple, context-aware searches to retrieve two classes of information for each property module: (1) direct data, which is any information directly linked to the queried material's exact structure, and (2) analog data, which comprises the properties of structurally analogous materials. These analogs are identified in real-time by launching a similarity search within the relevant data subset.

This \enquote{just-in-time} data augmentation is exceptionally powerful. If a queried material lacks data in a certain dimension—for instance, no experimental synthesis record exists—the platform can still provide crucial insights by displaying the synthesis conditions of its closest structural analogs. This mechanism effectively uses the collective knowledge of the entire database to enrich the profile of a single material, offering researchers valuable predictive hints and experimental starting points. It is this dynamic, similarity-driven approach that robustly bridges data gaps and transforms a collection of siloed datasets into an interconnected, intelligent knowledge base.

\begin{figure*}[t!]
  \centering
  \includegraphics[width=1.0\textwidth]{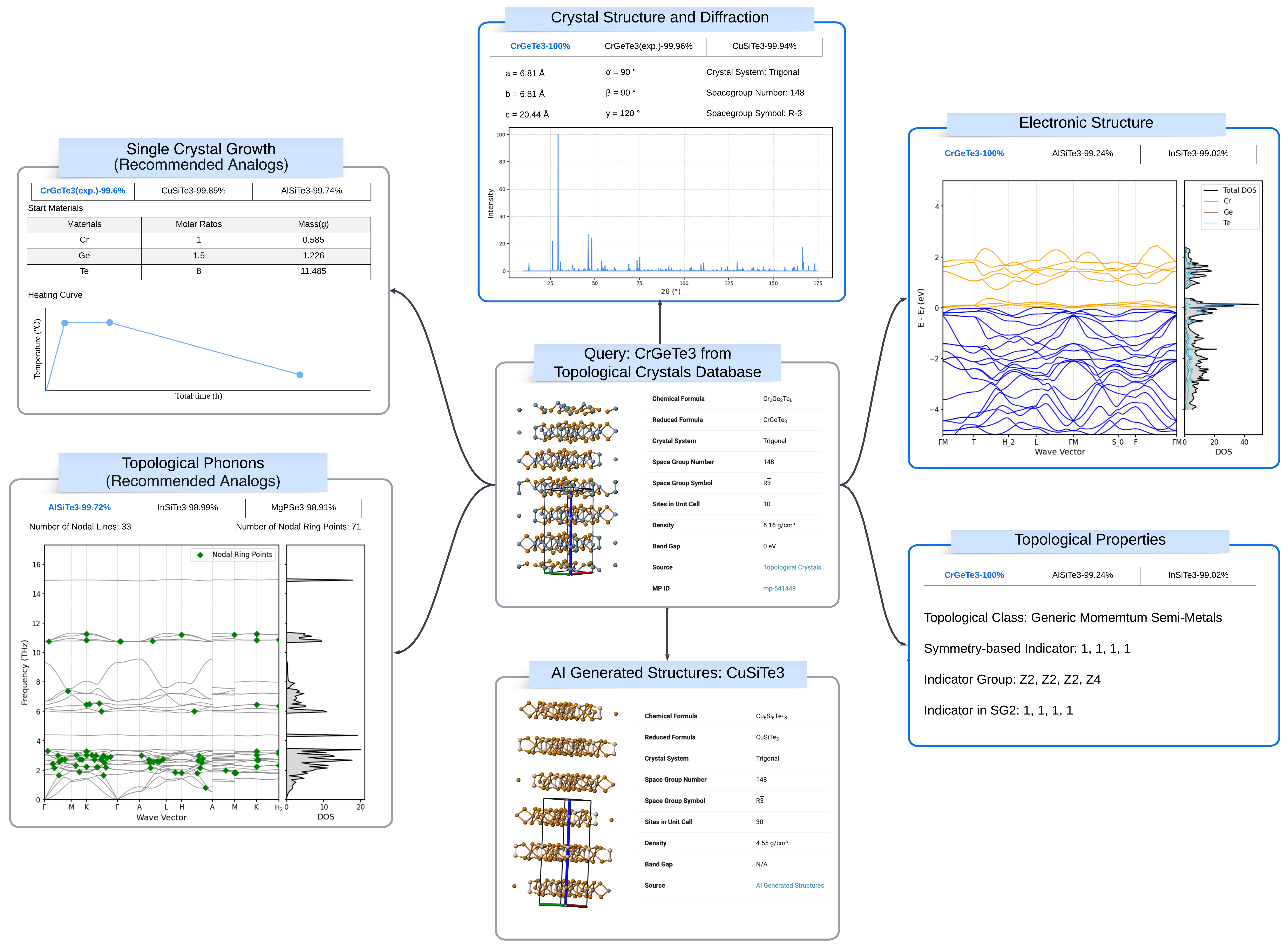}
  \caption[Data fusion workflow for \ch{CrGeTe3}.]{
    \textbf{Data fusion and discovery workflow for \ch{CrGeTe3}.}
    The platform's dual-axis analysis is triggered by a query for a target material.
    Horizontal Integration: Direct data for \ch{CrGeTe3} are aggregated across multiple modules (e.g., \enquote{Crystal Structure}, \enquote{Electronic Structure}) to build a deep, cross-modal profile linking experiment and theory.
    Vertical Comparison: The material profile is enriched with data from structural analogs. For modules where direct data is missing (e.g., \enquote{Single Crystal Growth}), the platform provides actionable references from known similar materials (e.g., \ch{AlSiTe3}). This comparison is further extended to novel, AI-generated structures (e.g., \ch{CuSiTe3}), enabling the exploration of uncharted chemical space for accelerated materials discovery.
  }
  \label{fig:similarity}
\end{figure*}

\subsection{Application example: Data fusion for \ch{CrGeTe3}}
\noindent
To illustrate the practical power of our structure-driven fusion mechanism, we present a case study on \ch{CrGeTe3}, a widely studied two-dimensional magnetic semiconductor\cite{sivadas_magnetic_2015,xu_interplay_2018,lin_tricritical_2017}. Data for this material are typically fragmented across our platform, making it an ideal example to demonstrate how our platform enables materials analysis along two orthogonal axes: horizontal integration for a single material and vertical comparison across similar materials. This dual-faceted workflow is visually encapsulated in Figure \ref{fig:similarity}.

First, the platform facilitates horizontal integration, creating a comprehensive, multi-modal profile for the target material itself. When a user queries \ch{CrGeTe3}, the system aggregates all its direct data by linking records from disparate sources. This process connects, for example, its experimental crystal structure from a diffraction database with its calculated electronic band structure from a topological database. This unified view enables powerful cross-modal analyses, such as correlating experimental conditions with theoretical properties. Even if a direct record is missing, this horizontal integration can fill a gap; for instance, if an experimental structure lacks a corresponding theoretical calculation, our similarity mechanism links it to the closest available computational entry, providing a robust theoretical proxy.

Building on this complete single-material profile, the platform enables vertical comparison, a powerful data-driven workflow for hypothesis generation and knowledge discovery. This process situates the target material within its broader \enquote{structural family} by identifying a cohort of its closest structural analogs via vector similarity. This vertical analysis pioneers a dual-pronged exploration strategy. It begins with knowledge extraction from known materials. By comparing \ch{CrGeTe3} against existing compounds in the database (e.g., \ch{FePSe3}, \ch{AlSiTe3}), researchers can systematically mine for structure-property relationships and identify trends, providing data-driven guidance to optimize experimental growth and reduce trial-and-error cycles.

More profoundly, this workflow accelerates the in silico discovery-synthesis loop by integrating AI-generated candidate structures. The vertical comparison extends beyond known materials into the vast, uncharted chemical space of theoretically plausible structures generated by our integrated deep generative models. By comparing the target material against these novel, yet-undiscovered candidates, researchers can identify promising new compositions that may exhibit superior functionalities. This process of generating theoretical candidates and comparing them against established materials provides a direct, actionable pathway for guiding future synthesis efforts towards the most promising frontiers, thereby embodying a key tenet of the data-driven materials discovery paradigm.

In summary, the \ch{CrGeTe3} case study demonstrates a paradigm shift from static data retrieval to a dynamic, multi-faceted research process. The horizontal integration offers unprecedented data depth for a single material, while the vertical comparison—spanning both known data and AI-generated possibilities—provides the data breadth required for true discovery. This powerful combination establishes a virtuous cycle: integrated data fuels AI models, which in turn generate new knowledge and propose novel materials, setting the stage for an accelerated, inverse design approach to materials science\cite{wang_inverse_2022}.

Additional application examples, including a topological phonon material (\ch{CoSi}) and a nonlinear optical material (\ch{LiNbO3}), are provided in the Supplementary Materials (Figure~\ref{fig:cosi_similarity} and Figure~\ref{fig:linbo3_similarity}).

\subsection{Platform features and functionality}
\noindent
Building on its core data fusion engine, the MaterialsGalaxy platform provides a multi-layered suite of features designed to maximize data accessibility, usability, and analytical power. The primary user entry point is a web-based portal offering a rich, interactive data exploration experience. Through this interface, users can perform complex queries using a combination of filters—such as chemical composition, space group symmetry, and calculated property ranges—and assess material properties through a suite of integrated visualization tools, including a 3D crystal structure viewer and interactive plots for electronic and phononic band structures.

To support the growing need for data-driven research and ensure interoperability, the platform is designed with the FAIR Guiding Principles as a cornerstone\cite{wilkinson_fair_2016}. Data are programmatically accessible through a well-documented RESTful API. This API, which adheres to the OpenAPI specification and is implemented using the FastAPI framework, allows for automated, large-scale data retrieval in a structured JSON format suitable for direct integration into machine learning workflows and custom research pipelines. Full documentation, including endpoint specifications and detailed usage examples for the API, is provided in the Supplementary Information. For bulk analysis, key datasets are also available for direct download.

A key distinguishing feature of MaterialsGalaxy is its seamless integration of state-of-the-art AI tools that operate directly on the fused data to accelerate the research cycle. These include our conversational agent, TopoChat, which is a specialized large language model for condensed matter physics\cite{xu_enhancing_2024}. For generative inverse design, the platform offers a modular framework supporting multiple distinct generative strategies, including the conditional variational autoencoder Con-CDVAE\cite{ye_concdvae_2024} and DiffCSP++\cite{jiao_space_2024}, a diffusion model that rigorously incorporates space group constraints. To bridge the experimental-computational gap, the platform also integrates PXRDGen\cite{li_powder_2025}, an end-to-end model for de novo crystal structure determination directly from powder X-ray diffraction (PXRD) patterns. Finally, to enable high-throughput virtual screening, a suite of machine learning models provides on-the-fly predictions for key properties like formation energy on any user-provided or AI-generated structure. These powerful, integrated AI capabilities collectively transform the platform from a simple data repository into a dynamic and interactive discovery environment.

\section{Discussion and Conclusion}
\noindent
The implementation of the MaterialsGalaxy platform demonstrates that systematic data standardization, when coupled with dynamic, structure-driven data association, can effectively dismantle the long-standing data silos in condensed matter physics. Our platform provides a powerful infrastructure to accelerate data-driven materials discovery, yet key challenges and opportunities for future enhancement remain.The foremost challenge is the scarcity of high-quality experimental data, which remains a bottleneck for the field and directly affects the reliability of both data fusion and downstream AI models\cite{kulik_using_2025,dunn_benchmarking_2020}. Another critical limitation arises from the dependence of our fusion scheme on structural similarity, which makes it sensitive to imperfections in experimental data, such as missing atoms or inaccurate atomic positions\cite{spek_checkcif_2020}.To enhance robustness, our future work will focus on developing more invariant structure representations under uncertainty\cite{goodall_predicting_2020,antunes_crystal_2024,zhu_wycryst_2024}, for instance by leveraging graph neural network embeddings\cite{xie_crystal_2018,chen_graph_2019} and exploring multi-modal fusion strategies that integrate complementary data modalities such as electronic band structures, X-ray diffraction (XRD) patterns, and other spectroscopic features\cite{moro_multimodal_2025}. Given the limited sample size within individual modalities, we further plan to investigate cross-modal alignment approaches, such as CLIP-based frameworks\cite{radford_learning_2021}, to better align heterogeneous data and enable scalable multi-modal representation learning.

\section{Methods}
\subsection{Data acquisition and standardization}
\noindent
The MaterialsGalaxy platform integrates data from multiple sources as detailed in the Results section, including public databases, internal experimental data, and data extracted from scientific literature. Prior to fusion, all data undergo a rigorous standardization pipeline. Crystal structures, primarily from CIF files\cite{hall_crystallographic_1991}, are parsed and validated using the pymatgen library\cite{ong_python_2013}. Standardized representations (including primitive and conventional cells) are generated according to IUCr conventions, with symmetry analysis handled by the underlying spglib library\cite{togo_spglib_2024}, ensuring consistent structural representation across all sources. Chemical formulas and calculated properties are also standardized, with metadata schemas (e.g., computational parameters) enforced in collaboration with domain experts to ensure data comparability. Automated scripts facilitate this efficient data ingestion and standardization process.

\subsection{Structure vectorization}
\noindent
To enable structure-based similarity search, each standardized crystal structure is converted into a fixed-length numerical vector (structural fingerprint). We employed the SiteStatsFingerprint featurizer from the matminer library\cite{ward_matminer_2018}. This method computes local atomic environment fingerprints and then calculates their statistics (mean and maximum) across all sites to generate a final, 122-dimensional structure-level vector. This representation effectively encodes key structural and chemical information into a format suitable for high-speed similarity comparison.
\subsection{Vector database and similarity search implementation}

\noindent
The generated structural fingerprint vectors are indexed for efficient retrieval using an implementation of the Approximate Nearest Neighbor (ANN) search algorithm. Specifically, we utilize an index based on the Hierarchical Navigable Small World (HNSW) graph method\cite{malkov_efficient_2020}, a state-of-the-art technique for high-dimensional vector search, often implemented in libraries such as Faiss\cite{johnson_billionscale_2021}. The index is constructed to balance memory usage and search accuracy. Similarity is quantified using the Cosine Similarity metric. Critically, data linking occurs dynamically at query time. A query structure's vector is used to retrieve its k-nearest neighbors (typically k=10-50) from the index within the relevant data subset. This ANN approach enables sub-second query times on datasets of millions of vectors, providing a scalable and responsive fusion experience.

\section*{Data Availability}
\noindent
The data supporting the findings of this study are publicly accessible through the Materials Galaxy platform and its associated services. The main web portal, available at \url{https://materialsgalaxy.iphy.ac.cn}, provides interactive browsing and visualization of all integrated data.

For programmatic access, a comprehensive, OpenAPI-compliant RESTful API is provided. The API offers tiered access: endpoints for searching and retrieving summary-level information for all materials are open to the public without authentication. Access to detailed data records for individual materials, such as full crystal structures and specific calculated properties, requires a free, user-registered API key. Full interactive documentation for the API is available at \url{https://materialsgalaxy.iphy.ac.cn/docs}, with usage guides at \url{https://materialsgalaxy.iphy.ac.cn/guides/api}.

To facilitate reproducibility and large-scale analysis, a static snapshot of the core data summary, containing essential information such as chemical formulas and crystal structures (in CIF format) for the materials presented in this study, is available for download on \url{https://materialsgalaxy.iphy.ac.cn/downloads}

The underlying datasets integrated into Materials Galaxy are hosted and maintained by the Condensed Matter Physics Data Center (CMPDC) (\url{https://cmpdc.iphy.ac.cn}). Raw data originating from external public databases (e.g., COD) are subject to their original licenses and remain available from their respective repositories.

\section*{Code Availability}
\noindent
The MaterialsGalaxy platform relies exclusively on publicly available, open-source libraries for its core scientific methodologies, as cited throughout the text. The key procedures for data standardization, structure vectorization, and similarity search can be reproduced using libraries such as Pymatgen \cite{ong_python_2013}, Matminer \cite{ward_matminer_2018}, and an appropriate ANN library implementing HNSW\cite{malkov_efficient_2020}. Further details are provided in the Methods section.

\section*{Acknowledgements}
\noindent
We wish to express our sincere gratitude to the researchers and their teams who contributed the foundational datasets that constitute MaterialsGalaxy. Specifically, we thank Shifeng Jin for the Crystal Structure and Diffraction data\cite{cmpdc_diffraction}; Youguo Shi for the Single Crystal Growth experimental data\cite{cmpdc_singlecrystal}; Jinbo Pan for the Layered Structures Database\cite{cmpdc_layered}; Qingbo Yan for the 2D Ferroelectric Materials Database\cite{cmpdc_ferroelectric}; Xingqiu Chen for the Topological Phonon Database\cite{synl_phonon}; and Yong Xu for the Nonlinear Optical Database\cite{cmpdc_nlo}. This work was supported by the Science Center of the National Natural Science Foundation of China (Grant No. 12188101), the National Natural Science Foundation of China (Grants No. 12274436 and No. 11921004), the National Key R\&D Program of China (Grants No. 2023YFA1607400 and No. 2022YFA1403800), H.W. acknowledges support from the New Cornerstone Science Foundation through the XPLORER PRIZE.
The AI-driven experiments, simulations and model training were performed on the robotic AI-Scientist platform of the Chinese Academy of Science.

\bibliographystyle{iopart-num}
\bibliography{references}

\providecommand{\newblock}{}
\begin{thebibliography}{10}
\expandafter\ifx\csname url\endcsname\relax
  \def\url#1{{\tt #1}}\fi
\expandafter\ifx\csname urlprefix\endcsname\relax\def\urlprefix{URL }\fi
\providecommand{\eprint}[2][]{\url{#2}}

\bibitem{zagorac_recent_2019}
Zagorac D, M{\"u}ller H, Ruehl S, Zagorac J and Rehme S 2019 {\em Journal of Applied Crystallography\/} {\bf 52} 918--925 ISSN 1600-5767

\bibitem{grazulis_crystallography_2009}
Gra{\v z}ulis S, Chateigner D, Downs R~T, Yokochi A~F~T, Quir{\'o}s M, Lutterotti L, Manakova E, Butkus J, Moeck P and Le~Bail A 2009 {\em Journal of Applied Crystallography\/} {\bf 42} 726--729 ISSN 0021-8898

\bibitem{grazulis_crystallography_2012}
Gra{\v z}ulis S, Da{\v s}kevi{\v c} A, Merkys A, Chateigner D, Lutterotti L, Quir{\'o}s M, Serebryanaya N~R, Moeck P, Downs R~T and Le~Bail A 2012 {\em Nucleic Acids Research\/} {\bf 40} D420--D427 ISSN 1362-4962, 0305-1048

\bibitem{groom_cambridge_2016}
Groom C~R, Bruno I~J, Lightfoot M~P and Ward S~C 2016 {\em Acta Crystallographica Section B: Structural Science, Crystal Engineering and Materials\/} {\bf 72} 171--179 ISSN 2052-5206

\bibitem{villars_pauling_2004}
Villars P, Berndt M, Brandenburg K, Cenzual K, Daams J, Hulliger F, Massalski T, Okamoto H, Osaki K, Prince A, Putz H and Iwata S 2004 {\em Journal of Alloys and Compounds\/} {\bf 367} 293--297 ISSN 0925-8388

\bibitem{jain_commentary_2013}
Jain A, Ong S~P, Hautier G, Chen W, Richards W~D, Dacek S, Cholia S, Gunter D, Skinner D, Ceder G and Persson K~A 2013 {\em APL Materials\/} {\bf 1}

\bibitem{curtarolo_aflow_2012}
Curtarolo S, Setyawan W, Hart G~L~W, Jahnatek M, Chepulskii R~V, Taylor R~H, Wang S, Xue J, Yang K, Levy O, Mehl M~J, Stokes H~T, Demchenko D~O and Morgan D 2012 {\em Computational Materials Science\/} {\bf 58} 218--226 ISSN 0927-0256

\bibitem{kirklin_open_2015}
Kirklin S, Saal J~E, Meredig B, Thompson A, Doak J~W, Aykol M, R{\"u}hl S and Wolverton C 2015 {\em npj Computational Materials\/} {\bf 1} 15010 ISSN 2057-3960

\bibitem{yang_matcloud_2018}
Yang X, Wang Z, Zhao X, Song J, Zhang M and Liu H 2018 {\em Computational Materials Science\/} {\bf 146} 319--333 ISSN 0927-0256

\bibitem{draxl_nomad_2019}
Draxl C and Scheffler M 2019 {\em Journal of Physics: Materials\/} {\bf 2} 036001 ISSN 2515-7639

\bibitem{talirz_materials_2020}
Talirz L, Kumbhar S, Passaro E, Yakutovich A~V, Granata V, Gargiulo F, Borelli M, Uhrin M, Huber S~P, Zoupanos S, Adorf C~S, Andersen C~W, Sch{\"u}tt O, Pignedoli C~A, Passerone D, VandeVondele J, Schulthess T~C, Smit B, Pizzi G and Marzari N 2020 {\em Scientific Data\/} {\bf 7} 299 ISSN 2052-4463

\bibitem{miao_atomly}
Miao L and Sheng M 2020 Atomly \urlprefix\url{https://atomly.net/}

\bibitem{hey_fourth_2009}
Hey T, Tansley S, Tolle K and Gray J 2009 {\em The {{Fourth Paradigm}}: {{Data-Intensive Scientific Discovery}}\/} (Microsoft Research) ISBN 978-0-9825442-0-4 \urlprefix\url{https://www.microsoft.com/en-us/research/publication/fourth-paradigm-data-intensive-scientific-discovery/}

\bibitem{agrawal_perspective_2016}
Agrawal A and Choudhary A 2016 {\em APL Materials\/} {\bf 4} 053208 ISSN 2166-532X

\bibitem{ramprasad_machine_2017}
Ramprasad R, Batra R, Pilania G, {Mannodi-Kanakkithodi} A and Kim C 2017 {\em npj Computational Materials\/} {\bf 3} 54 ISSN 2057-3960

\bibitem{lookman_active_2019}
Lookman T, Balachandran P~V, Xue D and Yuan R 2019 {\em npj Computational Materials\/} {\bf 5} 21 ISSN 2057-3960

\bibitem{schleder_dft_2019}
Schleder G~R, Padilha A~C~M, Acosta C~M, Costa M and Fazzio A 2019 {\em Journal of Physics: Materials\/} {\bf 2} 032001 ISSN 2515-7639

\bibitem{merchant_scaling_2023}
Merchant A, Batzner S, Schoenholz S~S, Aykol M, Cheon G and Cubuk E~D 2023 {\em Nature\/} {\bf 624} 80--85 ISSN 1476-4687

\bibitem{butler_machine_2018}
Butler K~T, Davies D~W, Cartwright H, Isayev O and Walsh A 2018 {\em Nature\/} {\bf 559} 547--555 ISSN 1476-4687

\bibitem{choudhary_recent_2022}
Choudhary K, DeCost B, Chen C, Jain A, Tavazza F, Cohn R, Park C~W, Choudhary A, Agrawal A, Billinge S~J~L, Holm E, Ong S~P and Wolverton C 2022 {\em npj Computational Materials\/} {\bf 8} 59 ISSN 2057-3960

\bibitem{zhong_explainable_2022}
Zhong X, Gallagher B, Liu S, Kailkhura B, Hiszpanski A and Han T~Y~J 2022 {\em npj Computational Materials\/} {\bf 8} 204 ISSN 2057-3960

\bibitem{vu_understanding_2023}
Vu T~S, Ha M~Q, Nguyen D~N, Nguyen V~C, Abe Y, Tran T, Tran H, Kino H, Miyake T, Tsuda K and Dam H~C 2023 {\em npj Computational Materials\/} {\bf 9} 215 ISSN 2057-3960

\bibitem{sanchez-lengeling_inverse_2018}
{Sanchez-Lengeling} B and {Aspuru-Guzik} A 2018 {\em Science\/} {\bf 361} 360--365

\bibitem{gubernatis_machine_2018}
Gubernatis J~E and Lookman T 2018 {\em Physical Review Materials\/} {\bf 2} 120301

\bibitem{ma_mlmd_2024}
Ma J, Cao B, Dong S, Tian Y, Wang M, Xiong J and Sun S 2024 {\em npj Computational Materials\/} {\bf 10} 59 ISSN 2057-3960

\bibitem{oganov_structure_2019}
Oganov A~R, Pickard C~J, Zhu Q and Needs R~J 2019 {\em Nature Reviews Materials\/} {\bf 4} 331--348 ISSN 2058-8437

\bibitem{chan_application_2022}
Chan C~H, Sun M and Huang B 2022 {\em EcoMat\/} {\bf 4} e12194 ISSN 2567-3173

\bibitem{griesemer_accelerating_2023}
Griesemer S~D, Xia Y and Wolverton C 2023 {\em Nature Computational Science\/} {\bf 3} 934--945 ISSN 2662-8457

\bibitem{stein_progress_2019}
Stein H~S and Gregoire J~M 2019 {\em Chemical Science\/} {\bf 10} 9640--9649 ISSN 2041-6539

\bibitem{pyzer-knapp_accelerating_2022}
{Pyzer-Knapp} E~O, Pitera J~W, Staar P~W~J, Takeda S, Laino T, Sanders D~P, Sexton J, Smith J~R and Curioni A 2022 {\em npj Computational Materials\/} {\bf 8} 84 ISSN 2057-3960

\bibitem{szymanski_autonomous_2023}
Szymanski N~J, Rendy B, Fei Y, Kumar R~E, He T, Milsted D, McDermott M~J, Gallant M, Cubuk E~D, Merchant A, Kim H, Jain A, Bartel C~J, Persson K, Zeng Y and Ceder G 2023 {\em Nature\/} {\bf 624} 86--91 ISSN 0028-0836, 1476-4687

\bibitem{kalidindi_materials_2015}
Kalidindi S~R and Graef M~D 2015 {\em Annual Review of Materials Research\/} {\bf 45} 171--193 ISSN 1531-7331, 1545-4118

\bibitem{himanen_datadriven_2019}
Himanen L, Geurts A, Foster A~S and Rinke P 2019 {\em Advanced Science\/} {\bf 6} 1900808 ISSN 2198-3844

\bibitem{andersen_optimade_2021}
Andersen C~W, Armiento R, Blokhin E, Conduit G~J, Dwaraknath S, Evans M~L, Fekete {\'A}, Gopakumar A, Gra{\v z}ulis S, Merkys A, Mohamed F, Oses C, Pizzi G, Rignanese G~M, Scheidgen M, Talirz L, Toher C, Winston D, Aversa R, Choudhary K, Colinet P, Curtarolo S, Di~Stefano D, Draxl C, Er S, Esters M, Fornari M, Giantomassi M, Govoni M, Hautier G, Hegde V, Horton M~K, Huck P, Huhs G, Hummelsh{\o}j J, Kariryaa A, Kozinsky B, Kumbhar S, Liu M, Marzari N, Morris A~J, Mostofi A~A, Persson K~A, Petretto G, Purcell T, Ricci F, Rose F, Scheffler M, Speckhard D, Uhrin M, Vaitkus A, Villars P, Waroquiers D, Wolverton C, Wu M and Yang X 2021 {\em Scientific Data\/} {\bf 8} 217 ISSN 2052-4463

\bibitem{tshitoyan_unsupervised_2019}
Tshitoyan V, Dagdelen J, Weston L, Dunn A, Rong Z, Kononova O, Persson K~A, Ceder G and Jain A 2019 {\em Nature\/} {\bf 571} 95--98 ISSN 1476-4687

\bibitem{gupta_matscibert_2022}
Gupta T, Zaki M, Krishnan N~M~A and Mausam 2022 {\em npj Computational Materials\/} {\bf 8} 102 ISSN 2057-3960

\bibitem{pyzer-knapp_foundation_2025}
{Pyzer-Knapp} E~O, Manica M, Staar P, Morin L, Ruch P, Laino T, Smith J~R and Curioni A 2025 {\em npj Computational Materials\/} {\bf 11} 61 ISSN 2057-3960

\bibitem{jiang_applications_2025}
Jiang X, Wang W, Tian S, Wang H, Lookman T and Su Y 2025 {\em npj Computational Materials\/} {\bf 11} 79 ISSN 2057-3960

\bibitem{green_fulfilling_2017}
Green M~L, Choi C~L, {Hattrick-Simpers} J~R, Joshi A~M, Takeuchi I, Barron S~C, Campo E, Chiang T, Empedocles S, Gregoire J~M, Kusne A~G, Martin J, Mehta A, Persson K, Trautt Z, Van~Duren J and Zakutayev A 2017 {\em Applied Physics Reviews\/} {\bf 4} 011105 ISSN 1931-9401

\bibitem{wu_universal_2024}
Wu Y, Wang C~F, Ju M~G, Jia Q, Zhou Q, Lu S, Gao X, Zhang Y and Wang J 2024 {\em Nature Communications\/} {\bf 15} 138 ISSN 2041-1723

\bibitem{raccuglia_machinelearningassisted_2016}
Raccuglia P, Elbert K~C, Adler P~D~F, Falk C, Wenny M~B, Mollo A, Zeller M, Friedler S~A, Schrier J and Norquist A~J 2016 {\em Nature\/} {\bf 533} 73--76 ISSN 1476-4687

\bibitem{zhang_topological_2009}
Zhang H, Liu C~X, Qi X~L, Dai X, Fang Z and Zhang S~C 2009 {\em Nature Physics\/} {\bf 5} 438--442 ISSN 1745-2481

\bibitem{weng_weyl_2015}
Weng H, Fang C, Fang Z, Bernevig B~A and Dai X 2015 {\em Physical Review X\/} {\bf 5} 011029

\bibitem{xie_crystal_2018}
Xie T and Grossman J~C 2018 {\em Physical Review Letters\/} {\bf 120} 145301

\bibitem{chen_graph_2019}
Chen C, Ye W, Zuo Y, Zheng C and Ong S~P 2019 {\em Chemistry of Materials\/} {\bf 31} 3564--3572 ISSN 0897-4756

\bibitem{schutt_schnet_2018}
Sch{\"u}tt K~T, Sauceda H~E, Kindermans P~J, Tkatchenko A and M{\"u}ller K~R 2018 {\em The Journal of Chemical Physics\/} {\bf 148} ISSN 0021-9606

\bibitem{ward_matminer_2018}
Ward L, Dunn A, Faghaninia A, Zimmermann N~E~R, Bajaj S, Wang Q, Montoya J, Chen J, Bystrom K, Dylla M, Chard K, Asta M, Persson K~A, Snyder G~J, Foster I and Jain A 2018 {\em Computational Materials Science\/} {\bf 152} 60--69 ISSN 0927-0256

\bibitem{xu_enhancing_2024}
Xu H, Zhang B, Jin Z, Zhu T, Wu Q and Weng H 2024 Enhancing {{Large Language Models}} with {{Domain-Specific Knowledge}}: {{The Case}} in {{Topological Materials}} (\textit{Preprint} \eprint{2409.13732})

\bibitem{ye_concdvae_2024}
Ye C~Y, Weng H~M and Wu Q~S 2024 {\em Computational Materials Today\/} {\bf 1} 100003 ISSN 2950-4635

\bibitem{iphy_matelab}
{Condensed Matter Physics Data Center, Institute of Physics, Chinese Academy of Sciences} {{MatElab}}: {{Electronic Laboratory}} for {{Material Science}} \urlprefix\url{https://matelab.iphy.ac.cn}

\bibitem{cmpdc_diffraction}
{Condensed Matter Physics Data Center, Institute of Physics, Chinese Academy of Sciences} Crystal {{Structure}} and {{Diffraction Database}} \urlprefix\url{https://cmpdc.iphy.ac.cn/diff/}

\bibitem{cmpdc_materiae}
{Condensed Matter Physics Data Center, Institute of Physics, Chinese Academy of Sciences} Materiae: {{Topological Materials Database}} \urlprefix\url{https://cmpdc.iphy.ac.cn/materiae/}

\bibitem{zhang_catalogue_2019}
Zhang T, Jiang Y, Song Z, Huang H, He Y, Fang Z, Weng H and Fang C 2019 {\em Nature\/} {\bf 566} 475--479 ISSN 1476-4687

\bibitem{li_computation_2021}
Li J, Liu J, Baronett S~A, Liu M, Wang L, Li R, Chen Y, Li D, Zhu Q and Chen X~Q 2021 {\em Nature Communications\/} {\bf 12} 1204 ISSN 2041-1723

\bibitem{synl_phonon}
{Condensed Matter Physics Data Center, Institute of Physics, Chinese Academy of Sciences} and {Institute of Metal Research, Chinese Academy of Sciences} Topological {{Phonon Database}} \urlprefix\url{http://www.phonon.synl.ac.cn/}

\bibitem{cmpdc_ferroelectric}
{Condensed Matter Physics Data Center, Institute of Physics, Chinese Academy of Sciences} and {Univeristy of Chinese Academy of Sciences} {{2D Ferroelectric Materials Database}} \urlprefix\url{https://cmpdc.iphy.ac.cn/fedb/}

\bibitem{cmpdc_nlo}
{Condensed Matter Physics Data Center, Institute of Physics, Chinese Academy of Sciences} and {Department of Physics, Tsinghua University} Nonlinear {{Optical Database}} \urlprefix\url{https://cmpdc.iphy.ac.cn/nlo/}

\bibitem{ong_python_2013}
Ong S~P, Richards W~D, Jain A, Hautier G, Kocher M, Cholia S, Gunter D, Chevrier V~L, Persson K~A and Ceder G 2013 {\em Computational Materials Science\/} {\bf 68} 314--319 ISSN 0927-0256

\bibitem{togo_spglib_2024}
Togo A, Shinohara K and Tanaka I 2024 {\em Science and Technology of Advanced Materials: Methods\/} {\bf 4} 2384822 ISSN null

\bibitem{kresse_efficient_1996}
Kresse G and Furthm{\"u}ller J 1996 {\em Physical Review B\/} {\bf 54} 11169--11186

\bibitem{malkov_efficient_2020}
Malkov Y~A and Yashunin D~A 2020 {\em IEEE Transactions on Pattern Analysis and Machine Intelligence\/} {\bf 42} 824--836 ISSN 0162-8828

\bibitem{sivadas_magnetic_2015}
Sivadas N, Daniels M~W, Swendsen R~H, Okamoto S and Xiao D 2015 {\em Physical Review B\/} {\bf 91} 235425

\bibitem{xu_interplay_2018}
Xu C, Feng J, Xiang H and Bellaiche L 2018 {\em npj Computational Materials\/} {\bf 4} 57 ISSN 2057-3960

\bibitem{lin_tricritical_2017}
Lin G~T, Zhuang H~L, Luo X, Liu B~J, Chen F~C, Yan J, Sun Y, Zhou J, Lu W~J, Tong P, Sheng Z~G, Qu Z, Song W~H, Zhu X~B and Sun Y~P 2017 {\em Physical Review B\/} {\bf 95} 245212

\bibitem{wang_inverse_2022}
Wang J, Wang Y and Chen Y 2022 {\em Materials\/} {\bf 15} 1811 ISSN 1996-1944

\bibitem{wilkinson_fair_2016}
Wilkinson M~D, Dumontier M, Aalbersberg I~J, Appleton G, Axton M, Baak A, Blomberg N, Boiten J~W, {da Silva Santos} L~B, Bourne P~E, Bouwman J, Brookes A~J, Clark T, Crosas M, Dillo I, Dumon O, Edmunds S, Evelo C~T, Finkers R, {Gonzalez-Beltran} A, Gray A~J~G, Groth P, Goble C, Grethe J~S, Heringa J, {'t Hoen} P~A~C, Hooft R, Kuhn T, Kok R, Kok J, Lusher S~J, Martone M~E, Mons A, Packer A~L, Persson B, {Rocca-Serra} P, Roos M, {van Schaik} R, Sansone S~A, Schultes E, Sengstag T, Slater T, Strawn G, Swertz M~A, Thompson M, {van der Lei} J, {van Mulligen} E, Velterop J, Waagmeester A, Wittenburg P, Wolstencroft K, Zhao J and Mons B 2016 {\em Scientific Data\/} {\bf 3} 160018 ISSN 2052-4463

\bibitem{jiao_space_2024}
Jiao R, Huang W, Liu Y, Zhao D and Liu Y 2024 Space {{Group Constrained Crystal Generation}} (\textit{Preprint} \eprint{2402.03992})

\bibitem{li_powder_2025}
Li Q, Jiao R, Wu L, Zhu T, Huang W, Jin S, Liu Y, Weng H and Chen X 2025 {\em Nature Communications\/} {\bf 16} 7428 ISSN 2041-1723

\bibitem{kulik_using_2025}
Kulik H~J 2025 {\em Journal of Materials Research\/} {\bf 40} 833--848 ISSN 2044-5326

\bibitem{dunn_benchmarking_2020}
Dunn A, Wang Q, Ganose A, Dopp D and Jain A 2020 {\em npj Computational Materials\/} {\bf 6} 138 ISSN 2057-3960

\bibitem{spek_checkcif_2020}
Spek A~L 2020 {\em Acta Crystallographica Section E: Crystallographic Communications\/} {\bf 76} 1--11 ISSN 2056-9890

\bibitem{goodall_predicting_2020}
Goodall R~E~A and Lee A~A 2020 {\em Nature Communications\/} {\bf 11} 6280 ISSN 2041-1723

\bibitem{antunes_crystal_2024}
Antunes L~M, Butler K~T and {Grau-Crespo} R 2024 {\em Nature Communications\/} {\bf 15} 10570 ISSN 2041-1723

\bibitem{zhu_wycryst_2024}
Zhu R, Nong W, Yamazaki S and Hippalgaonkar K 2024 {\em Matter\/} {\bf 7} 3469--3488 ISSN 2590-2385

\bibitem{moro_multimodal_2025}
Moro V, Loh C, Dangovski R, Ghorashi A, Ma A, Chen Z, Kim S, Lu P~Y, Christensen T and Solja{\v c}i{\'c} M 2025 {\em Newton\/} {\bf 1} ISSN 2950-6360

\bibitem{radford_learning_2021}
Radford A, Kim J~W, Hallacy C, Ramesh A, Goh G, Agarwal S, Sastry G, Askell A, Mishkin P, Clark J, Krueger G and Sutskever I 2021 Learning {{Transferable Visual Models From Natural Language Supervision}} (\textit{Preprint} \eprint{2103.00020})

\bibitem{hall_crystallographic_1991}
Hall S~R, Allen F~H and Brown I~D 1991 {\em Acta Crystallographica Section A\/} {\bf 47} 655--685 ISSN 1600-5724

\bibitem{johnson_billionscale_2021}
Johnson J, Douze M and J{\'e}gou H 2021 {\em IEEE Transactions on Big Data\/} {\bf 7} 535--547 ISSN 2332-7790

\bibitem{cmpdc_singlecrystal}
{Condensed Matter Physics Data Center, Institute of Physics, Chinese Academy of Sciences} Single {{Crystal Growth Database}} \urlprefix\url{https://cmpdc.iphy.ac.cn/mlab/}

\bibitem{cmpdc_layered}
{Condensed Matter Physics Data Center, Institute of Physics, Chinese Academy of Sciences} Layered {{Materials Database}} \urlprefix\url{https://cmpdc.iphy.ac.cn/layered/}

\end{thebibliography}

\clearpage
\clearpage
\appendix
\onecolumn

\section{Application Example: Topological Phonon Material CoSi}
\label{sec:si_cosi}

Beyond the \ch{CrGeTe3} case study presented in the main text, we further demonstrate the versatility of the MaterialsGalaxy platform using cubic cobalt silicide (\ch{CoSi}), a prototypical topological phonon material. 
Topological phonons—lattice vibrations with nontrivial band topology—have recently emerged as a frontier topic in condensed matter physics, with potential applications in phononic and quantum devices.

\begin{figure}[H]
  \centering
  \includegraphics[width=\textwidth]{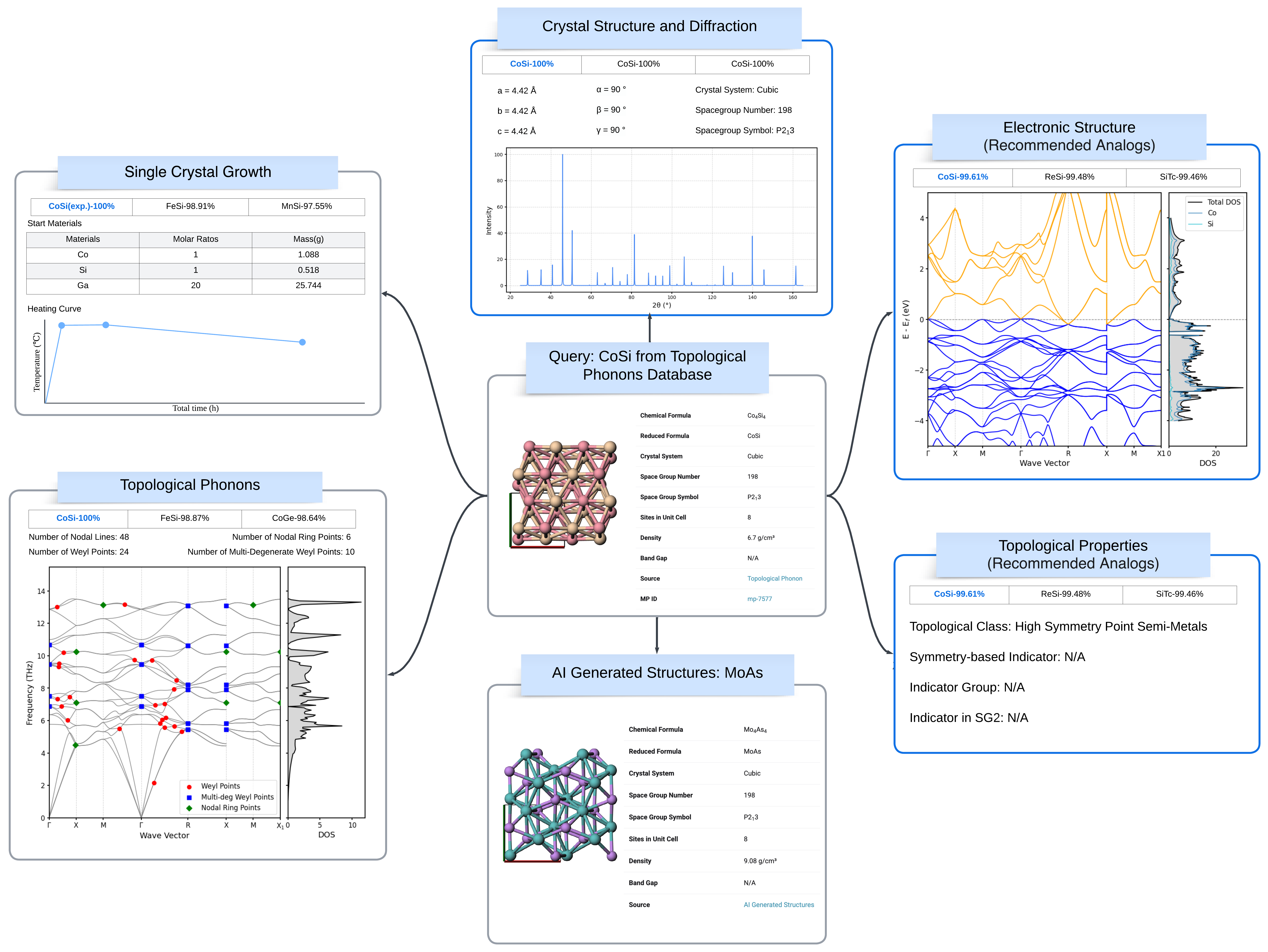}
  \caption{Integrated data visualization for \ch{CoSi}, a topological phonon material. 
  Horizontal integration aggregates multi-modal data for \ch{CoSi}, including crystal structure and diffraction patterns, electronic band structure, topological classification, phononic dispersion, and consolidated single-crystal growth records from multiple experiments. 
  Vertical comparison identifies structurally similar materials for both single-crystal growth and topological phonon properties, alongside AI-generated candidate structures (\ch{MoAs}), demonstrating the platform's capability to connect experimental synthesis data, theoretical calculations, and computational predictions in a unified framework.}
  \label{fig:cosi_similarity}
\end{figure}

Figure~\ref{fig:cosi_similarity} illustrates how the integrated data ecosystem enables comprehensive exploration of \ch{CoSi}. 
Through horizontal integration, the platform aggregates multi-modal data for this material, including its crystal structure and diffraction data, electronic band structure, topological properties, and phononic properties. 
Notably, the platform's single-crystal growth module contains a wealth of experimental synthesis records for \ch{CoSi}; one representative record is displayed in this visualization. 
These extensive growth datasets provide invaluable practical guidance for researchers seeking to reproduce or optimize synthesis conditions.

Through vertical comparison, the platform demonstrates its structural-similarity-driven search capabilities. 
The system identifies structurally analogous compounds across all property modules that may exhibit related characteristics. 
Additionally, the platform integrates AI-generated candidate structures (e.g., \ch{MoAs} shown in the figure), extending the exploration beyond experimentally known materials into theoretically predicted chemical space.

Overall, the \ch{CoSi} example underscores how MaterialsGalaxy unifies multi-source data, AI-assisted similarity analysis, and generative design into a coherent workflow, demonstrating the platform's broad applicability across diverse functional material systems.

\clearpage

\section{Application Example: Nonlinear Optical Material \texorpdfstring{\ch{LiNbO3}}{LiNbO3}}
\label{sec:si_linbo3}

To further demonstrate the versatility of the MaterialsGalaxy platform, we showcase lithium niobate (\ch{LiNbO3}), a prototypical nonlinear optical (NLO) and ferroelectric material. 
\ch{LiNbO3} is one of the most widely studied optical crystals, featuring strong polarization, large electro-optic coefficients, and a pronounced nonlinear optical response. 
Its rich experimental and computational datasets make it an ideal system for demonstrating the integration of structure, property, and optical-response data.

\begin{figure}[H]
  \centering
  \includegraphics[width=\textwidth]{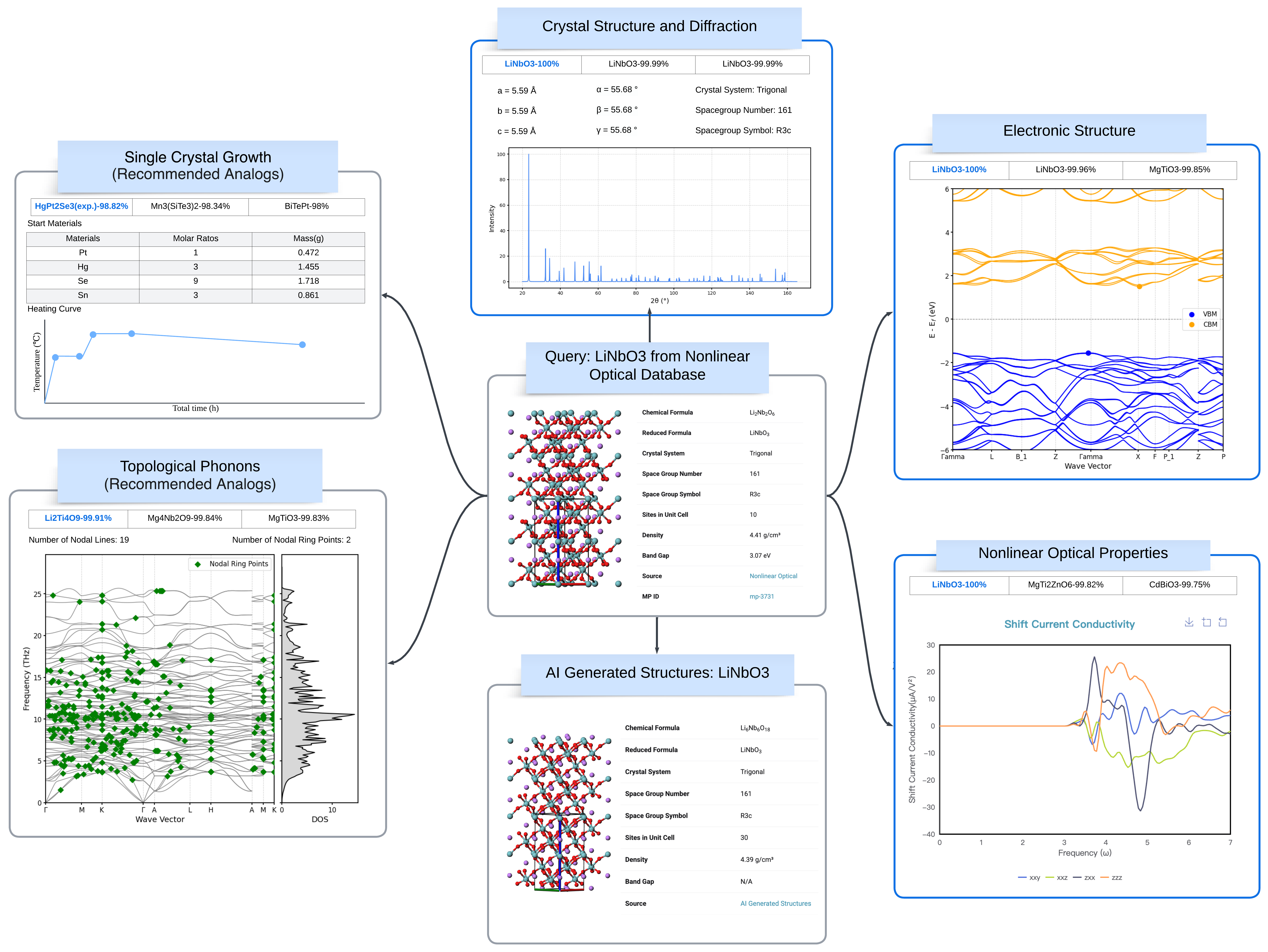}
  \caption{Integrated data visualization for \ch{LiNbO3}, a benchmark nonlinear optical material. 
  Horizontal integration summarizes experimental and theoretical data including crystal growth, electronic, and optical properties. 
  Vertical comparison lists structurally similar compounds identified through vector-based similarity search, enabling comparative analysis within the Li–Nb–O materials family.}
  \label{fig:linbo3_similarity}
\end{figure}

Figure~\ref{fig:linbo3_similarity} illustrates how the integrated data ecosystem enables comprehensive exploration of \ch{LiNbO3}. 
Through horizontal integration, the platform aggregates multi-modal data for this material, including its crystal structure and diffraction data, electronic band structure, and calculated nonlinear optical response properties.

Through vertical comparison, the platform demonstrates its structural-similarity-driven search capabilities. 
The system identifies structurally analogous compounds across all property modules that may exhibit related characteristics, providing complementary data to fill gaps in missing modalities.

Overall, this example highlights how MaterialsGalaxy connects theoretical and experimental data within a unified, data-driven framework for materials exploration and discovery.

\clearpage

\section*{Supplementary Information: MaterialsGalaxy API}
\label{sec:si_api}

\section{Overview and Adherence to FAIR Principles}
The MaterialsGalaxy platform provides a comprehensive, high-performance RESTful Application Programming Interface (API) to enable programmatic access to all its integrated data. The API is designed with the \textbf{FAIR Guiding Principles} (Findable, Accessible, Interoperable, and Reusable) as a core tenet, facilitating advanced data-driven research and integration into automated workflows.

\begin{itemize}
  \item \textbf{Findable}: Each material entry is assigned a unique, persistent identifier (\texttt{mg\_id}). The API provides powerful search endpoints that allow materials to be found based on a rich set of compositional, structural, and property-based metadata.
  \item \textbf{Accessible}: The API is accessible via the standard HTTPS protocol. Publicly available summary data can be retrieved without authentication, while access to detailed, non-public data is managed through a secure bearer token authentication system, ensuring controlled and traceable access.
  \item \textbf{Interoperable}: The API strictly adheres to the \textbf{OpenAPI specification}, providing a machine-readable contract for all endpoints. Data is exchanged in the standardized and widely adopted \textbf{JSON} format. The backend is built on a Pydantic-enforced data schema, guaranteeing that all data payloads are well-structured and consistent.
  \item \textbf{Reusable}: The rich metadata returned by the API provides essential context (e.g., data source, calculated properties) that is crucial for the proper reuse of data in new research contexts and for training machine learning models.
\end{itemize}

The base URL for all API v1 endpoints is:
\begin{verbatim}
https://materialsgalaxy.iphy.ac.cn/api/v1/
\end{verbatim}

Complete, interactive API documentation is available at: \url{https://materialsgalaxy.iphy.ac.cn/docs}.

\section{Authentication}
Certain endpoints, particularly those providing detailed crystal structures or specific calculated properties, require authentication. Users can obtain a personal API key from their account profile page on the main platform portal. The key must be included in the request header as a Bearer Token:
\begin{verbatim}
Authorization: Bearer YOUR_API_KEY_HERE
\end{verbatim}

\section{Core Endpoint Examples}
The following examples demonstrate common usage patterns for the API. Python examples use the \texttt{requests} library.

\subsection{Example 1: Retrieving a Specific Material Summary}
This endpoint retrieves the core summary information for a material given its unique \texttt{mg\_id}. This endpoint is public and does not require authentication.

\begin{itemize}
  \item \textbf{Endpoint}: \texttt{GET /materials/summary/{mg\_id}}
  \item \textbf{Python Example}:
        \begin{verbatim}
import requests

API_ROOT = "https://materialsgalaxy.iphy.ac.cn/api/v1"
MG_ID = "mg-1"

response = requests.get(f"{API_ROOT}/materials/summary/{MG_ID}")
if response.status_code == 200:
    data = response.json()
    print(data)
else:
    print(f"Error: {response.status_code}")
\end{verbatim}
\end{itemize}

\subsection{Example 2: Advanced Granular Search for Materials}
The API supports powerful, multi-parameter searches for materials. This example demonstrates a search for materials containing both Silicon (Si) and Oxygen (O).

\begin{itemize}
  \item \textbf{Endpoint}: \texttt{GET /materials/summary}
  \item \textbf{Description}: This endpoint accepts numerous query parameters for filtering materials. Common parameters include compositional filters (\texttt{elements}, \texttt{formula}), structural filters (\texttt{crystal\_systems}, \texttt{spacegroups}), and property ranges (\texttt{band\_gap\_min}).
  \item \textbf{Python Example}:
        \begin{verbatim}
import requests

API_ROOT = "https://materialsgalaxy.iphy.ac.cn/api/v1"
params = {
    "elements": "Si,O",
    "page": 1,
    "page_size": 20
}
response = requests.get(f"{API_ROOT}/materials/summary", params=params)
if response.status_code == 200:
    data = response.json()
    print(f"Found {data['total']} materials.")
    for material in data['data']:
        print(f"- {material['mg_id']}: {material['reduced_formula']}")
else:
    print(f"Error: {response.status_code}")
\end{verbatim}
\end{itemize}

\subsection{Example 3: Finding Structurally Similar Materials (Vector Search)}
This endpoint leverages the platform's core vector similarity search to find materials with similar crystal structures. This endpoint requires authentication.

\begin{itemize}
  \item \textbf{Endpoint}: \texttt{GET /materials/similarity}
  \item \textbf{Description}: Given a source material's \texttt{mg\_id}, this returns a ranked list of its closest structural analogs. The search can be filtered to specific property domains (e.g., \texttt{singleCrystalGrowth}).
  \item \textbf{Python Example}:
        \begin{verbatim}
import requests

API_ROOT = "https://materialsgalaxy.iphy.ac.cn/api/v1"
API_KEY = "YOUR_API_KEY_HERE"
headers = {"Authorization": f"Bearer {API_KEY}"}

params = {
    "mg_id": "mg-1",          # The material to find analogs for
    "property": "electronicStructure", # Context for the search
    "k": 5                    # Number of analogs to return
}
response = requests.get(f"{API_ROOT}/materials/similarity", params=params, headers=headers)

if response.status_code == 200:
    analogs = response.json()
    print("Found structural analogs:")
    for analog in analogs:
        print(f"- {analog['mg_id']}: {analog['reduced_formula']} "
              f"(Similarity Score: {analog['distance']:.4f})")
else:
    print(f"Error: {response.status_code}")
\end{verbatim}
\end{itemize}

\end{document}